\documentclass{elsart}
\usepackage{natbib}
\usepackage{graphicx}
\usepackage{amsmath}
\begin{document}
\begin{frontmatter}

\title{The dark energy--dominated Universe}
\author{Jos\'{e} Carlos N. de Araujo}
\address{Instituto Nacional de Pesquisas Espaciais - Divis\~ao de
Astrof\'\i sica \\ Av. dos Astronautas 1758, S\~ao Jos\'e dos
Campos, 12227-010 SP, Brazil}
\ead{jcarlos@das.inpe.br}

\begin{abstract} 
In this paper we investigate the epochs in which the Universe
started accelerating and when it began to become dark
energy--dominated (i.e., the dynamics of the expansion of the
Universe dominated by the dark energy). We provide analytic
expressions to calculate the redshifts of these epochs as a
function of density parameters. Moreover, we review and discuss
cosmological models with a dark energy component, which can have
an interesting characteristic, namely, they never stop
accelerating. This holds even if the Universe is at present time
either flat, open, or closed. If the dark energy is the
cosmological constant the Universe will eventually end up
undergoing an exponentially expansion phase, and the total density
parameter converging to $\Omega=1$. This is exactly what is
considered in inflationary scenario to generate the initial
conditions for the big bang. One can then argue that the Universe
begun with an inflationary phase and will end up with another
inflationary phase. Thus, it follows that in both the early and
the late Universe $\Omega \rightarrow 1$. We also discuss the
above issues in the context of the XCDM parametrization.
\end{abstract}






\end{frontmatter}

\maketitle

\section{Introduction}

There are many pieces of evidence for the existence of a component of the
Universe other than the baryonic matter and the non baryonic dark
matter (see, e.g., \citealp{peebles}).

This other component of the Universe could be Einstein's
cosmological constant, or a component that varies slowly with time
and space. Many authors name this component as dark energy or
quintessence, should it be a cosmological constant or a component
that acts like it.

From the measurements of the cosmic microwave anisotropy, the WMAP
satellite team has recently announced that there is convincing
evidence that most of the energy in the Universe today is dark.
This dark energy is gravitationally repulsive and accelerates the
expansion of the Universe (see, e.g., \citealp{spergel}), and
could well be Einstein's cosmological constant.

It is worth mentioning that earlier evidence for an accelerating
Universe was found by other astrophysical projects; The Supernova
Cosmology Project is an example (see, e.g., \citealp{perlmutter}).

Also, it is worth recalling that the existence of a dark energy
component is required for the age of the Universe be compatible
with the oldest objects of the Universe (see, e.g.,
\citealp{carrol}, amongst others, for a detailed discussion).

Although implicity~in~previous~papers~(see,~e.g.,
\citealp{vishwakarma,sahni,carrol,madsen,felten}, among others),
and~also in some textbooks (see, e.g., \citealp{peacock}), there
is still some room to discuss some interesting questions not yet
addressed to the same extent in the literature. Firstly, the epoch
when the Universe begins accelerating. Secondly, the epoch when
the Universe begins to be dark energy dominated. Thirdly, even
Universe models spatially closed can expand forever. Fourthly,
besides expanding forever spatially open, and closed models behave
like flat models for a sufficiently long time; this very fact
leads to an interesting conclusion, namely: the Universe begins
almost flat and after being open, flat, or closed at present time,
it will be flat in the future.

Since the dynamics is dominated by the dark energy, in the case of
a cosmological constant, the scale factor increases exponentially
for the late Universe. An interesting fact in such a case is that
the scale factor of the Universe at large time resembles the scale
factor of the inflationary era. One could think of the Universe as
having two inflationary eras, one at early times, and the other
one after several Hubble times. Instead, if the dark energy is in
the form of another accelerating component, the scale factor
increases as a power law, as for example in the XCDM model.

The plan of the present paper is the following. In
Section~\ref{sec-2} we discuss how the dark energy accelerates the
Universe and the epoch in which the Universe becomes dark energy
dominated; in Section~\ref{sec-3} we discuss the flatness of the
early and late Universe; in Section~\ref{sec-4} we consider a
putative late inflationary epoch; and finally in
Section~\ref{conc} we present the conclusions.

\section{The accelerating and the dark energy--dominated Universe}
\label{sec-2}
In this section we deal with the role of the dark
energy in accelerating the Universe. A useful quantity here is the
decelerating parameter, $q(a)$, namely
\begin{equation}\label{q}
q(a) \equiv - \frac{\ddot{a} a}{\dot{a}^{2}},
\end{equation}
\noindent (see, e.g., \citealp{peacock}) where the dot stands for
time derivatives, and $a$ ($\equiv R/R_{0}$; scale factor over its
present value) is the normalized scale factor. We refer the reader
to Appendix A, where one finds a series of useful equations used
in the present paper.

There are in the literature many different alternatives for the
dark energy (see, e.g., \citealp{peebles}), the most usual one is
the cosmological constant, $\Lambda$. In the present study we
restrict our attention to the dark energy in the so called XCDM
parametrization, which includes, as a particular case, the
cosmological constant.

In the XCDM parametrization the pressure, $p_{X}$, is written as
\begin{equation}\label{px}
p_{X}=w_{X}\rho_{X},
\end{equation}
\noindent where $w_{X}$ is a constant and $\rho_{X}$ is the energy
density of the XCDM fluid. The above equation for $p_{X}$ is also
known as the cosmic equation of state.

From the local energy conservation one finds that
\begin{equation}\label{rhoXp}
\rho_{X}\propto a^{{-3(1+w_{X})}}
\end{equation}
\noindent (see Appendix A for details).

Note that $w_{x}=-1$ implies that $\rho_{X}$ is constant. This is
nothing but a fluid representation of the cosmological constant.
If $w_{X}<-1$, $\rho_{X}$ is an increasing function of $a$. On the
other hand, if $ w_{X} > -1$, $\rho_{X}$ is a decreasing function
of $a$.

Note that the WMAP satellite team (see, e.g, \citealp{spergel})
reported that $w_{X} < -0.78$, therefore consistent with a
cosmological constant.

From the Friedmann equation it is easy to show that $q(a)$ can be
written in terms of the density parameters, namely
\begin{equation}\label{qa}
q(a)=\frac{1}{2}\;\Omega_{m}(a)+\Omega_{r}(a)+
\frac{1+3w_{X}}{2}\;\Omega_{X}(a);
\end{equation}
\par\noindent which can be written in terms of the present values
of density parameters -- $\Omega_{m}$, $\Omega_{r}$ and
$\Omega_{X}$, for matter, radiation and dark energy, respectively
-- namely
\begin{equation}\label{qaa}
q(a)=\frac{\Omega_{m} a
+2\Omega_{r}+(1+3w_{X})\;\Omega_{X}a^{1-3w_{X}}}{2E(a)},
\end{equation}
\par\noindent where $E(a)$ is found in Appendix A. Note that we
have represented the present values of the density parameters
simply by ``$\Omega$'s" and the corresponding subscripts.
Hereafter, any density parameter thus represented means its
present value, i.e., $\Omega_{i}\equiv\Omega_{i}(a=1)$.

\begin{figure}
\includegraphics[width=84mm]{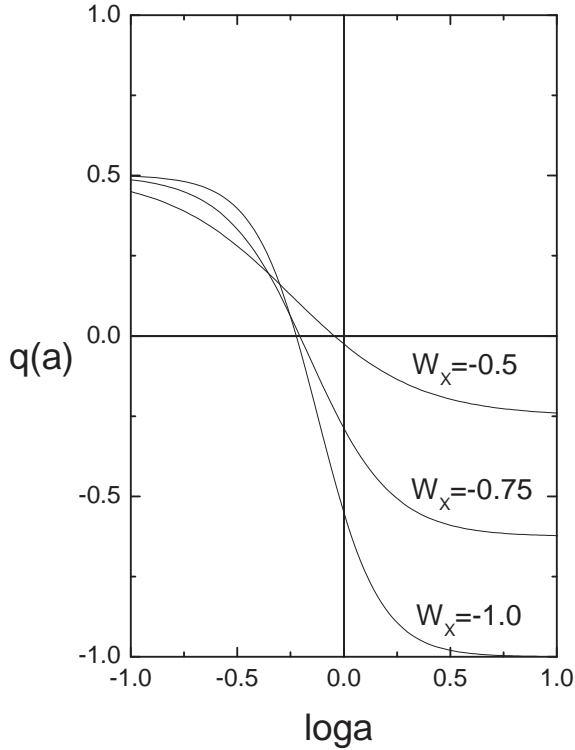}
\caption{Evolution of $q$ as a function of $a$ for a flat Universe
model with $\Omega_{m}=0.30$ and $\Omega_{X}=0.70$, for different
values of $w_{X}$.} \label{fig1}
\end{figure}

Note that for an accelerating Universe, i.e., $\ddot{a}>0$, it
means that $q(a)<0$. From the above equation one sees that if
\begin{eqnarray}
w_{X} < -\frac{1}{3} \nonumber ,
\end{eqnarray}
\noindent it may occur that
\begin{eqnarray}
q(a) < 0.    \nonumber
\end{eqnarray}
From the equations for the evolution of density parameters as a
function of the scale factor -- found in Appendix A, for example
--, one obtains a simple formula either for the redshift,
$z_{acc}$, or the scale factor, $a_{acc}$, in which the Universe
begins its accelerating epoch, namely
\begin{equation}\label{aacc}
a_{acc}=\left(\frac{-1}{1+3w_{X}}\,\frac{\Omega_{m}}
{\Omega_{X}}\right)^{-1/3w_{X}},
\end{equation}
\par and
\begin{equation}\label{zacc}
z_{acc}=\left[-(1+3w_{X})
\frac{\Omega_{X}}{\Omega_{m}}\right]^{-1/3w_{X}}-1.
\end{equation}
In Fig.~\ref{fig1} some illustrative examples are given on how $q$
varies as a function of the scale factor, $a$, for a flat Universe
model with $\Omega_{m}=0.30$ and $\Omega_{X}=0.70$, for different
values of $w_{X}$.  The lower $w_{X}$ is, the lower the redshift
in which the Universe begins to accelerate. Note that, for
$w_{X}=-1$ the acceleration of the Universe takes place at
$z_{acc}=0.67$.

\par It is worth mentioning that once the Universe has started
accelerating it keeps accelerating forever. The decelerating
parameter goes asymptotically to a finite value, namely
\begin{eqnarray}
q(a)  \rightarrow \frac{1+3w_{X}}{2} \;\;\;\;\;\;\; {\rm for} \;\;
w_{X} < - \frac{1}{3} \;\; {\rm and} \;\; a \gg 1 .\nonumber
\end{eqnarray}
Now, we show that the epoch in which the Universe becomes dark
energy dominated is not the epoch in which it starts accelerating.

The dark energy--dominated Universe begins when $\Omega_{X}
> \Omega_{m}$. From the equations for the evolution of the
density parameters one obtains the redshift (the scale factor)
$z_{eq}^{\ast}$ ($a_{eq}^{\ast}$) of matter--dark energy equality,
namely,
\begin{equation}\label{aeq}
a_{eq}^{\ast} = \left(\frac{\Omega_{m}}
{\Omega_{X}}\right)^{-1/3w_{X}},
\end{equation}
\par and
\begin{equation}\label{zeq}
z_{eq}^{\ast} =\left(
\frac{\Omega_{X}}{\Omega_{m}}\right)^{-1/3w_{X}}-1.
\end{equation}
Note that there is no the factor of $[-(1+3w_{X})]$ in the above
equations as compared to Eqs. \ref{aacc} and \ref{zacc}.

The reason why the accelerating and the dark energy--dominated
eras begin at different redshifts has to do with the following.
The decelerating parameter comes from the Friedman equation,
therefore the pressure and the energy density are implicitly taken
into account in the calculation of $z_{acc}$ ($a_{acc}$). On the
other hand, $z_{eq}^{\ast}$ ($a_{eq}^{\ast}$) is obtained by just
equating $\Omega_{X}$ with $\Omega_{m}$, involving therefore only
energy densities.

For a flat Universe model with $\Omega_{m}=0.30$ and
$\Omega_{X}=0.70$ for $w_{X}=-1$ the dark energy--dominated era
begins at $z_{eq}^{\ast}=0.33$, which is almost a factor of two
lower than $z_{acc}$ for this very model.

As it is well known, coincidentally, the accelerating and the dark
energy--dominated era took place quite recently.

\section{Flat in the beginning and at the end}
\label{sec-3}
\begin{figure}
\includegraphics[width=84mm]{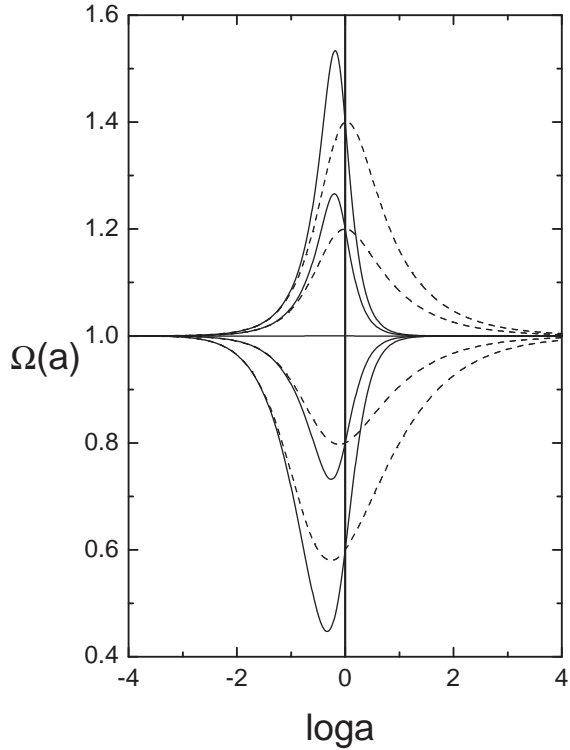}
\caption{A comparison of the evolution of different Universe
models: $\Omega=0.6$, 0.8, 1.0, 1.2 and 1.4, with
$\Omega_{X}~=~0.5$, 0.6, 0.7, 0.8 and 0.9, respectively, for
$w_{X}=-1$ (cosmological constant), solid lines, and $w_{X}=-0.5$,
dashed lines. Note that all of them are nearly flat in the
beginning of the Universe and in the future.} \label{fig2}
\end{figure}

\begin{figure}
\includegraphics[width=84mm]{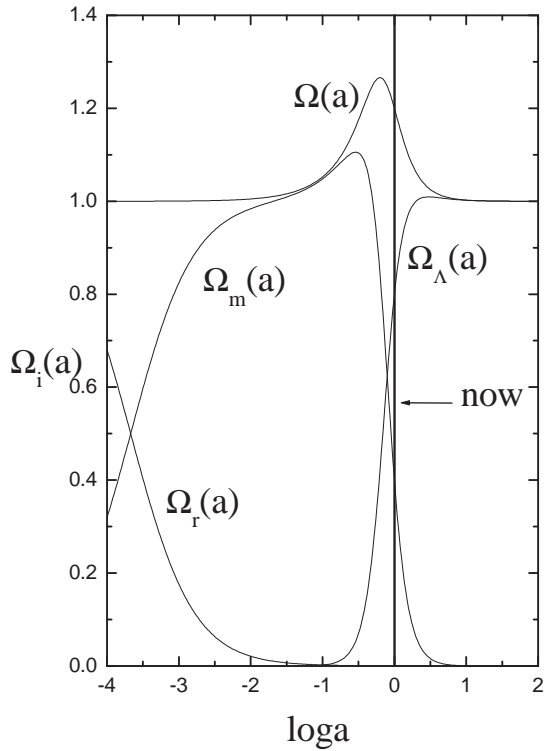}
\caption{The evolution of the density parameters as a function of
the scale factor for a Universe model with $\Omega=1.2$, where
$\Omega_{m}=0.4$ and $\Omega_{X}=0.8$. Note that, since we are
considering $w_{X}=-1$ (cosmological constant) the label of the
dark energy in the figure is $\Omega_{\Lambda}$ instead of
$\Omega_{X}$. Note that $\Omega_{i}(a)$ stands for the different
components of the Universe.} \label{fig3}
\end{figure}

\begin{figure}
\includegraphics[width=84mm]{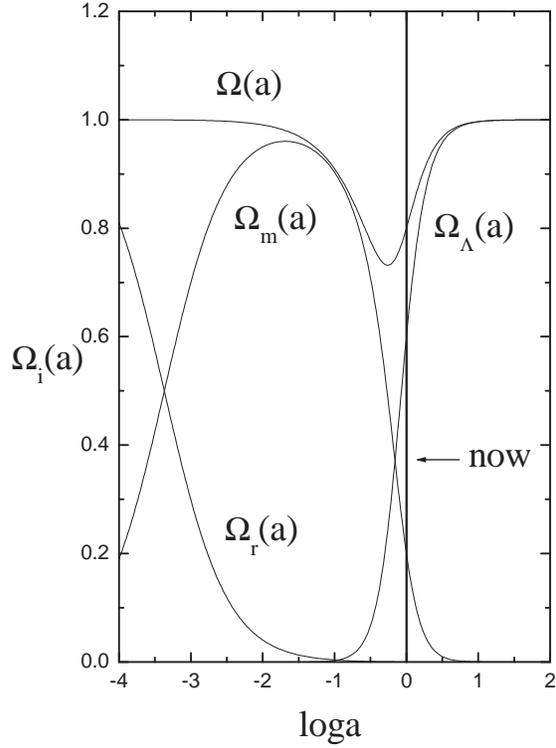}
\caption{The same as in Fig. 3 for a Universe model with
$\Omega=0.8$, where $\Omega_{m}=0.2$ and $\Omega_{\Lambda}=0.6$.}
\label{fig4}
\end{figure}

For models with $0<\Omega_{m}<1$ and $0<\Omega_{\Lambda}<1$ (where
$\Omega_{\Lambda}$ is the density parameter for the cosmological
contant), even for those combinations for which
$\Omega_{m}+\Omega_{\Lambda}>1$ the Universe expands forever (see,
e.g., \citealp{carrol,peacock}). It is worth stressing that for
these range of values of the density parameters, this effect has
to do with the fact that the energy density related to the
cosmological constant is constant during the expansion of the
Universe which keeps the Universe expanding forever.

Fig.~\ref{fig2} shows some examples on how the total density
parameter,
\begin{equation}
\Omega(a) = \Omega_{m}(a)+\Omega_{r}(a)+\Omega_{X}(a),
\end{equation}
\par\noindent evolves as a function of the scale factor, for
different models, namely: $\Omega=0.6$, 0.8, 1.0, 1.2 and 1.4,
with $\Omega_{X}=0.5$, 0.6, 0.7, 0.8 and 0.9, respectively, for
$w_{X}=-1$ (cosmological constant), solid lines, and $w_{X}=-0.5$,
dashed lines. Note that all of them are nearly flat in the
beginning of the Universe and in the future, where the dark energy
term accounts for this latter evolution. Obviously, most models
present in Fig.~\ref{fig2} are merely illustrative, since they are
not consistent with the present knowledge one has for the
cosmological parameters.

As can be seen in Appendix A, Eq. 10 can be written as follows
\begin{equation}\label{omegata}
\Omega(a)=\frac{\Omega_{r}+\Omega_{m}a+\Omega_{X}a^{1-3w_{X}}}
{E(a)},
\end{equation}
\noindent where $E(a)$ is also found in Appendix A.

\par Note that for small values of $a$, as is well known, the energy
density of the Universe is dominated by the radiation, then
\begin{eqnarray}
\Omega(a) \rightarrow \Omega_{r}(a) \rightarrow 1 \;\;\;\;\;\;\;
{\rm for} \;\; a \rightarrow 0. \nonumber
\end{eqnarray}
\par Also, the Universe at early time tends to be flat
independent of the present values of density parameters.

\par On the other hand, for large values of $a$ the energy density
is dominated by the dark energy, then
\begin{eqnarray}
\Omega(a) \rightarrow \Omega_{X}(a) \rightarrow 1 \;\;\;\;\;\;\;
{\rm for} \;\; w_{X} < - \frac{1}{3} \;\; {\rm and} \;\; a \gg 1.
\nonumber
\end{eqnarray}
\par Moreover, in the same way as that which occurs at early times,
the Universe behaves like a flat model, independent of the present
values of the density parameters.

In general, any quintessence model, which yields an energy density
that evolves with a power law in the scale factor with an exponent
greater than $-2$, has a similar behavior as above.

Therefore, the Universe begins indefinitely flat and will be in
the future indefinitely flat, even being either open or closed at
intermediate times.

To illustrate how the density parameters of the different
components of the Universe evolves, as a function of the scale
factor, we present two examples in Figs.~\ref{fig3} and
\ref{fig4}.

Fig.~\ref{fig3} refers to a closed model with $\Omega=1.2$, where
$\Omega_{m}=0.4$ and $\Omega_{X}=0.8$. In this figure $\Omega_{i}$
stands for the different components of the Universe. Note that,
since we are considering $w_{X}=-1$ (cosmological constant) the
label of the dark energy in the figure is $\Omega_{\Lambda}$
instead of $\Omega_{X}$.

Fig.~\ref{fig3} shows, as is already well known, that in the early
Universe the dynamics is dominated by the radiation (the
radiation--dominated era), after that, at $z\sim 10^{4}$, up to
recent times, the dynamics is dominated by the matter (the
matter--dominated era). Moreover, due to the existence of a dark
energy component in the Universe, there is the dark
energy--dominated era, where the dynamics is dominated by the dark
energy. This era begins at a recent time and dominates the
dynamics of the Universe indefinitely.

Fig.~\ref{fig4} shows the same as in Fig.~\ref{fig3} now for an
open model with a cosmological constant with $\Omega=0.8$, where
$\Omega_{m}=0.2$ and $\Omega_{\Lambda}=0.6$. The conclusions
concerning the evolution of this Universe model are the same as
the closed model discussed above.

\section{Is there a late inflationary epoch?}
\label{sec-4}

Let us now consider how the time evolution of the scale factor is
for the XCDM cosmology adopted here, to see, in particular, how
the evolution of the Universe is during the dark energy-dominated
era.

Note that the time evolution for the scale factor for the
radiation-dominated era is not affected by the dark energy. The
matter-dominated era is only affected at the end of this era, when
the contribution of the dark energy begins to be significant.

The time evolution of the scale factor, for several Hubble times,
is completely dominated by the dark energy, and is given by
\begin{eqnarray}\label{ats}
a = \left\{\begin{array}{ll}
t^{2/[3(1+w_{x})]} \quad &  {\rm for} \quad -1 < w_{x}\leq -1/3 \\ \\
\exp\sqrt{\frac{8\pi G}{3}\rho_{x}}t  \quad &  {\rm for} \quad  w_{x}=-1 \\
\end{array}\right.
\end{eqnarray}
Note that for $w_{x}=-1$, i.e., for a cosmological constant, the
scale factor evolves exponentially. This behavior is exactly the
same as that which occurs in inflationary cosmology to generate
the initial conditions for the big bang.

One could think of the Universe as having two inflationary eras,
one at early times and other one at large times, which would cause
the Universe to be indefinitely flat at early times and at future
times. The difference of this putative late inflation as compared
to the early inflation is that the former never stops.

Recall that the de Sitter model is devoid of matter or radiation,
there is only the contribution of a cosmological constant, and
therefore the Universe expands exponentially, as shown in Eq.
(12).

Since for several Hubble times there is no contribution of matter
for the dynamics of the expansion of the Universe in the XCDM
model considered here, because $\Omega_{m}\rightarrow 0 $, this
expansion phase of the dark energy-dominated era is the de Sitter
one.

On the other hand, in particular for $-1<w_{x}\leq -1/3$, the
Universe expands forever as a power law in the scale factor, and
$\Omega \rightarrow 1$ for several Hubble times. Again, one could
think of this as being an inflationary phase, but now as a power
law one.

It is worth noting that the behavior of the XCDM models studied
here is given, for several Hubble times, by Eq. (12), irrespective
of the Universe at present time being open, flat or closed.

In XCDM accelerating models the Universe never stop accelerating
unless the dark energy is such that the value of $w_{X}$ is not a
constant and is less then $- 1/3$ in some epoch of the evolution
of the Universe.

It is worth mentioning, however, that the dark energy, or the
quintessence, could well have a more complex behavior in such a
way that it causes an accelerating phase around the present time,
and for future times the Universe stops accelerating. This issue
is not discussed here and we refer the reader to the literature
for details.

\section{Conclusions}
\label{conc} In this paper we have discussed the role of the dark
energy in the evolution of the Universe. We have adopted, in
particular, the XCDM parametrization, where the cosmological
constant is a particular case of dark energy.

The epochs in which the Universe begun its accelerating phase and
when the Universe becomes dark energy--dominated have been
studied. In particular, expressions have been derived to calculate
the redshifts corresponding to these very epochs, in terms of
density parameters of a given Universe model.

Both the early and the late evolutions of the Universe have been
addressed. In particular, it has been shown that, for a broad
range of density parameters, the Universe expands forever, even
though at present time the Universe is closed. The dark energy is
reponsible for the Universe being asymptotically flat in the
future times.

If the dark energy is in the form of a cosmological constant the
Universe will expand exponentially at future times, as in the de
Sitter model. This resembles the inflationary era of the early
Universe. Thus, one can think of as the Universe having two
inflationary eras, one at early times and the other one at future
times.

For dark energy in which $-1 <w_{X}\leq -1/3$ the Universe expands
as a power law, becoming asymptotically flat.

\ack{I would like to thank Dr. Oswaldo Duarte Miranda for the
critical reading of the paper. The author has received financial
support from CNPq (grants 304666/02-5). I would also like to thank
the referee for the careful reading of the paper.}

\appendix
\section{Density parameters and related issues}

In this appendix we consider a series of equations which are
useful in various sections of the present article.

First, the Roberton-Walker metric reads
\begin{equation}\label{rw}
ds^{2}=c^{2}dt^{2}-R^{2}(t)dl^{2},
\end{equation}
\noindent where $dl^{2}$ depends only on the spatial coordinates,
$c$ is the velocity of light, $t$ is the universal time, and
$R(t)$ is the scale factor.

A useful quantity is the normalized scale factor, $a(t)$, which is
related to the scale factor as follows:
\begin{equation}\label{at}
a(t)=\frac{R(t)}{R_{0}},
\end{equation}
\noindent where $R_{0}$ is the present value of the scale factor,
which can be taken to be equal to unity.

Concerning the cosmological parameters it is implicit in many
textbooks and papers that one is referring to their present
values. It is well known, however, that the density parameters
evolve with the expansion of the Universe.

Whenever the density parameters appear in the present paper
without any functional dependence, it means that we are referring
to their present values.

Let us look at how the density parameters evolve in an expanding
Universe. Recall that the density parameter is defined as follows:
\begin{equation}\label{denpar}
\Omega_{i}(a)\equiv \frac{\rho_{i}(a)}{\rho_{c}(a)},
\end{equation}
\par\noindent where $\rho_{i}(a)$ is the energy density of \emph{i}th
component i.e., baryons, photons, dark matter, and dark energy,
for example as a function of the scale factor $a$, and
$\rho_{c}(a)$ is the critical density as a function of $a$.

Recalling that the Friedmann equation reads
\begin{equation}\label{FE}
\dot{R}^{2}-\frac{8\pi G}{3}\;\rho R^{2}=-kc^{2},
\end{equation}
\noindent where $k$ is either -1, 0 or +1, for open, flat and
closed Universes, respectively. Adopting $k=0$ one can obtain from
the above equation an expression for the density, which is called
critical density, i.e., the density of a flat Universe.

Thus, the critical density reads
\begin{equation}\label{dp}
\rho_{c}(a)=\frac{3H^{2}}{8\pi G},
\end{equation}
and
\begin{equation}\label{H}
H=\frac{\dot{R}}{R}
\end{equation}
\noindent is the Hubble parameter, and the ``dot" stands for time
derivatives.

\par It is worth mentioning that instead of following the time evolution
of the density parameters we follow, throughout the paper, their
evolution either as function of the scale factor ($a$ or $R$) or
as function of the redshift, z. These very quantities are related
to each other as follows:
\begin{eqnarray}
a = \frac{R}{R_{0}}= \frac{1}{1+z}. \nonumber
\end{eqnarray}
The energy content of the Universe is divided into
pressureless matter, in the form of baryonic matter and non
baryonic weakly interacting cold dark matter, radiation, and dark
energy.

The matter and the radiation energy densities depend on the scale
factor as follows
\begin{equation}\label{rhom}
\rho_{m}\propto R^{-3},
\end{equation}
\begin{equation}\label{rhor}
\rho_{r}\propto R^{-4},
\end{equation}
For the dark energy, recall that in the XCDM parametrization, the
pressure is written
\begin{equation}\label{apx}
p_{X}=w_{X}\rho_{X}.
\end{equation}
From the local energy conservation, namely
\begin{equation}\label{ec}
\dot{\rho}=-3\;H(t)\;(\rho+p),
\end{equation}
\noindent it follows that
\begin{equation}\label{arhoXp}
\dot{\rho_{X}}=-3\frac{\dot{a}}{a}\rho_{X}(1+w_{X}),
\end{equation}
\noindent thus
\begin{equation}\label{rhoX}
\rho_{X}\propto a^{-3(1+w_{X})}
\end{equation}
Now, from the Friedmann equation one obtains
\begin{equation}\label{omega}
\frac{kc^{2}}{H^{2}R^{2}}=\Omega_{m}(a)+\Omega_{r}(a)+\Omega_{X}(a)
-1
\end{equation}
Defining the curvature density parameter as
\begin{equation}\label{omegak}
\Omega_{k}(a)\equiv-\frac{kc^{2}}{H^{2}R^{2}},
\end{equation}
\noindent it follows that
\begin{equation}\label{omegas}
\Omega_{m}(a)+\Omega_{r}(a)+\Omega_{X}(a)+\Omega_{k}(a)=1;
\end{equation}
\noindent this means that if $\Omega_{k}(a)$ is added to the other
density parameters one gets a unity value.

In the present paper, $\Omega$ without a subscript denotes the
total density parameter without the curvature term, namely
\begin{equation}\label{omegat}
\Omega(a) = \Omega_{m}(a)+\Omega_{r}(a)+\Omega_{X}(a)
\end{equation}
Now, we present how the density parameters evolve as function of
$a$. It is easy to show that
\begin{equation}\label{8pigrho}
\frac{8\pi G\rho}{3}=H^{2}_{0}\left(\Omega_{X}a^{-3(1+w_{X})}
+\Omega_{m}a^{-3}+\Omega_{r}a^{-4}\right).
\end{equation}
Using equation (\ref{omega}) one obtains
\begin{equation}\label{Hubble}
H^{2}(a)=H^{2}_{0}\left[\Omega_{X}a^{-3(1+w_{X})}+\Omega_{k}a^{-2}
+\Omega_{m}a^{-3}+\Omega_{r}a^{-4} \right].
\end{equation}
Finally, using the definition of density parameter for each
component of the Universe one obtains
\begin{equation}\label{omegara}
\Omega_{r}(a)=\frac{\Omega_{r}}{E(a)},
\end{equation}
\begin{equation}\label{omegama}
\Omega_{m}(a)=\frac{\Omega_{m}}{E(a)}\;a,
\end{equation}
\begin{equation}\label{omegaka}
\Omega_{k}(a)=\frac{\Omega_{k}}{E(a)}\;a^{2},
\end{equation}
\begin{equation}\label{omegaxa}
\Omega_{X}(a)=\frac{\Omega_{X}}{E(a)}\;a^{1-3w_{X}},
\end{equation}
\begin{equation}
\Omega(a)=\frac{\Omega_{r}+\Omega_{m}a+\Omega_{X}a^{1-3w_{X}}}
{E(a)};
\end{equation}
\noindent where
\begin{equation}
E(a)=\Omega_{r}+\Omega_{m}a+\Omega_{k}a^{2}+\Omega_{X}a^{1-3w_{X}}.
\end{equation}

\begin{thebibliography}{}
%
\bibitem {carrol} Carroll, S. M., Press, W. P., \& Turner, E. L. 1992, ARAA, 30, 599
%
\bibitem {felten} Felten, J. E., \& Isaacman, R. 1986, Rev. Mod. Phys., 58, 689
%
\bibitem {madsen} Madsen, M. S., Mimoso, J. P., Butcher, J. A., \& Ellis, G. F. R.
1992, PRD, 46, 1399
%
\bibitem {peacock} Peacock, J. 1999, Cosmological Physics (Cambridge University Press)
%
\bibitem {peebles} Peebles, P. J. E. 2003, Rev. Mod. Phys.,
75, 559
%
\bibitem {perlmutter} Perlmutter, S. et al. 1999, AJ, 517, 565
%
\bibitem {sahni} Sahni, V., \& Starobinsky, A. 2000,
Int. J. Mod. Phys. D, 9, 373
%
\bibitem {spergel} Spergel, D. N., Verde, L., Peiris, H. V.,
et al. 2003, ApJS, 148, 175
%
\bibitem {vishwakarma} Vishwakarma, R. G. 2002, MNRAS, 331, 776
%
\end{thebibliography}
\end{document}